\documentclass[12pt]{article}

\usepackage{amsmath}
\usepackage{graphicx}
\usepackage{cite}

\newcommand{\ket}[1]{| #1 \rangle}
\newcommand{\bra}[1]{\langle #1 |}
\newcommand{\hcs}[1]{#1^\dagger #1}
\newcommand{\expv}[1]{\langle #1 \rangle}

\begin{document}

\begin{center}
{\Large\bf Allowed region and optimal measurement \\
for information versus disturbance
in quantum measurements}
\vskip .6 cm
Hiroaki Terashima
\vskip .4 cm
{\it Department of Physics, Faculty of Education, Gunma University, \\
Maebashi, Gunma 371-8510, Japan}
\vskip .6 cm
\end{center}

\begin{abstract}
We present graphs of information versus disturbance
for general quantum measurements of completely unknown states.
Each piece of information and disturbance is quantified by two measures:
(i) the Shannon entropy and estimation fidelity for the information
and (ii) the operation fidelity and physical reversibility for the disturbance.
These measures are calculated for a single outcome
and are plotted on four types of information--disturbance planes
to show their allowed regions.
In addition, we discuss the graphs of
these metrics averaged over all possible outcomes
and the optimal measurements when
saturating the upper bounds on the information for a given disturbance.
The results considerably broaden the perspective of trade-offs
between information and disturbances in quantum measurements.
\end{abstract}

\begin{flushleft}
{\footnotesize
}
\end{flushleft}

\section{Introduction}
In quantum theory,
a measurement that provides information about a system
inevitably disturbs the state of the system,
unless the original state is a
classical mixture of the eigenstates of an observable.
This feature is not only of great interest
to the foundations of quantum mechanics
but also plays an important role in quantum information processing
and communication~\cite{NieChu00},
such as in quantum cryptography~\cite{BenBra84,Ekert91,Bennet92,BeBrMe92}.
As a result, the relationship between information and disturbances
has been the subject of numerous
studies~\cite{FucPer96,Banasz01,FucJac01,BanDev01,Barnum02,%
DArian03,Ozawa04,GenPar05,MiFiFi05,Maccon06,Sacchi06,BusSac06,Banasz06,%
BuHaHo07,RenFan14,FGNZ15,ShKuUe16}
over many years.
Most studies have only discussed the disturbance
in terms of the size of the state change.
However, the disturbance can also be discussed in terms of the reversibility of
the state change~\cite{Terash10,Terash11,CheLee12,Terash15}
because the state change can be recovered
with a nonzero probability of success
if the measurement is physically reversible~\cite{UedKit92,UeImNa96,Ueda97}.

Intuitively, if a measurement provides more information about a system,
the measurement changes the state of the system
by a greater degree and the change becomes more irreversible.
To show this trade-off,
various inequalities have been derived using different formulations.
For example, Banaszek~\cite{Banasz01} derived
an inequality between the amount of information gain and
the size of the state change using two fidelities,
and Cheong and Lee~\cite{CheLee12} derived
an inequality between the amount of information gain and
the reversibility of the state change
using the fidelity and reversal probability.
These inequalities have been
verified~\cite{SRDFM06,BaChKi08,CZXTLX14,LRHLK14}
in single-photon experiments.

In this paper,
we present graphs of information versus disturbance
for general quantum measurements
of a $d$-level system in a completely unknown state.
The information is quantified
by the Shannon entropy~\cite{FucPer96}
and the estimation fidelity~\cite{Banasz01},
whereas the disturbance is quantified
by the operation fidelity~\cite{Banasz01}
and the physical reversibility~\cite{KoaUed99}.
These metrics are calculated for a single outcome
using the general formulas derived in Ref.~\cite{Terash15}
and are plotted on four types of information--disturbance planes
to show the allowed regions.
Moreover, we show the allowed regions for
these metrics averaged over all possible outcomes
via an analogy with the center of mass.
The allowed regions explain the structure of
the relationship between the information and disturbance
including both the upper and lower bounds on the information
for a given disturbance,
even though the lower bounds can be violated by
non-quantum effects
such as classical noise and the observer's non-optimal estimation.
In particular, optimal measurements saturating the upper bounds
are shown to be different for
the four types of information--disturbance pairs.
Therefore, our results broaden our understanding of
the effects of quantum measurements
and provide a useful tool for quantum information processing
and communication.

Two of the above bounds have been shown
by Banaszek~\cite{Banasz01} and Cheong and Lee~\cite{CheLee12}
to be inequalities for the average values
via different methods than ours.
The most important difference is that
they directly discussed
the information and disturbance averaged over outcomes,
whereas we start with those pertaining to each single outcome
derived~\cite{Terash15} in the context of
a physically reversible measurement~\cite{UedKit92,UeImNa96,Ueda97}.
Even though trade-offs between
information and disturbance are conventionally discussed
using the average
values~\cite{FucPer96,Banasz01,BanDev01,Barnum02,Sacchi06,Banasz06},
physically reversible measurements
strongly imply trade-offs at the level of a single outcome~\cite{DArian03}.
That is, in a physically reversible measurement,
whenever a second measurement called the reversing measurement
recovers the pre-measurement state of the first measurement,
it erases all the information obtained by the first measurement
(see the Erratum of Ref.~\cite{Royer94}).
This state recovery with information erasure occurs not on average
but only when the reversing measurement
yields a preferred single outcome.

Moreover, starting from the level of a single outcome
greatly simplifies the derivation of the allowed regions
and optimal measurements.
It is easy to show the allowed regions
pertaining to a single outcome
because the information and disturbance
pertaining to a single outcome contain only
a definite number of bounded parameters
and have some useful invariances under parameter transformations.
From these allowed regions,
the allowed regions for the average values
are shown using a graphical method based on an analogy with the center of mass,
which makes it easy to construct the optimal measurements.
In fact, without our method,
it would be difficult
to find all of the bounds and optimal measurements.

The rest of this paper is organized as follows.
Section~\ref{sec:formulation} reviews the procedure
for quantifying the information and disturbances in quantum measurements.
Sections~\ref{sec:region} and \ref{sec:average} show
the allowed regions for information and disturbance
pertaining to a single outcome
and those for the average values over all possible outcomes.
Section~\ref{sec:optimal} discusses
the optimal measurements to show their differences
for the four types of information--disturbance pairs.
Section~\ref{sec:summary} summarizes our results.

\section{\label{sec:formulation}Information and Disturbance}
First, the amount of information
provided by a measurement is quantified.
Suppose that the $d$-level system to be measured
is known to be in one of a set of
predefined pure states $\{\ket{\psi(a)}\}$.
The probability for $\ket{\psi(a)}$ is given by $p(a)$;
however, which $\ket{\psi(a)}$ is actually assigned
to the system is unknown.
Here we focus on the case where 
no prior information concerning the system is available,
assuming that $\{\ket{\psi(a)}\}$ is
a set of all the possible pure states and that $p(a)$ is uniform
according to the normalized invariant measure over the pure states.
Because $\{\ket{\psi(a)}\}$ in this case is a continuous set of states,
the index $a$ actually represents a set of continuous parameters
such as the hyperspherical coordinates in $2d$ dimensions
as in Ref.~\cite{Terash15},
where the summation over $a$ is replaced with
an integral over the coordinates
using the hyperspherical volume element.

It is measured to obtain information about the state of the system.
A quantum measurement can be described by a set of
measurement operators $\{\hat{M}_m\}$~\cite{NieChu00} that satisfy
\begin{equation}
\sum_m\hcs{\hat{M}_m}=\hat{I},
\label{eq:completeness}
\end{equation}
where $m$ denotes the outcome of the measurement
and $\hat{I}$ is the identity operator.
Here, the quantum measurement has been assumed to be
ideal~\cite{NieCav97} or efficient~\cite{FucJac01}
in the sense that it does not have classical noise
yielding mixed post-measurement states
because we focus on the quantum nature of the measurement.
When the system is in a state $\ket{\psi(a)}$,
the measurement $\{\hat{M}_m\}$ yields an outcome $m$ with probability
\begin{equation}
p(m|a)=\bra{\psi(a)}\hcs{\hat{M}_m}\ket{\psi(a)},
\end{equation}
changing the state into
\begin{equation}
\ket{\psi(m,a)}=\frac{1}{\sqrt{p(m|a)}}\,\hat{M}_m\ket{\psi(a)}.
\label{eq:postMeasurement}
\end{equation}
Each measurement operator
can be decomposed by a singular-value decomposition, such as
\begin{equation}
\hat{M}_m=\hat{U}_m\hat{D}_m\hat{V}_m,
\end{equation}
where $\hat{U}_m$ and $\hat{V}_m$ are unitary operators
and $\hat{D}_m$ is a diagonal operator
in an orthonormal basis $\{\ket{i}\}$ with $i=1,2,\ldots,d$ such that
\begin{equation}
\hat{D}_m=\sum_{i} \lambda_{mi} \ket{i}\bra{i}.
\end{equation}
The diagonal elements $\{\lambda_{mi}\}$ are called
the singular values of $\hat{M}_m$ and satisfy $0\le\lambda_{mi}\le1$.

From the outcome $m$, the state of the system can be partially deduced.
For example, Bayes's rule states that, given an outcome $m$,
the probability that the state was $\ket{\psi(a)}$
is given by
\begin{equation}
 p(a|m) =\frac{p(m|a)\,p(a)}{p(m)},
\end{equation}
where $p(m)$ is the total probability of outcome $m$,
\begin{equation}
 p(m) =\sum_a p(m|a)\,p(a).
\end{equation}
That is, the outcome $m$ changes the probability distribution
for the states from $\{p(a)\}$ to $\{p(a|m)\}$.
This change decreases the Shannon entropy,
which is known as a measure of the lack of information:
\begin{align}
  I(m)  &=\left[-\sum_a p(a)\log_2 p(a)\right] \notag \\
        &   \qquad{}-\left[-\sum_a p(a|m)\log_2 p(a|m)\right].
\label{eq:defInformation}
\end{align}
Therefore, $I(m)$, which we define as the information gain,
quantifies the amount of information
provided by the outcome $m$ of
the measurement $\{\hat{M}_m\}$~\cite{DArian03,TerUed07b}
and is explicitly written in terms of
the singular values of $\hat{M}_m$ as~\cite{Terash15}
\begin{align}
 I(m) &=\log_2d-\frac{1}{\ln2}\Bigl[\eta(d)- 1\Bigr] \notag \\
    & \qquad {}-\log_2\sigma_m^2  +\frac{1}{\sigma_m^2}
    \sum_{i}\frac{\lambda_{mi}^{2d}\log_2\lambda_{mi}^2}
      {\prod_{k\neq i}\left( \lambda_{mi}^2-\lambda_{mk}^2\right)},
\label{eq:information}
\end{align}
where
\begin{equation}
  \eta(n) =\sum^{n}_{k=1}\frac{1}{k}, \qquad
  \sigma_m^2 =\sum_{i}\lambda_{mi}^2.
\end{equation}
Note that $I(m)$ satisfies
\begin{equation}
 0\le I(m) \le\log_2d-\frac{1}{\ln2}[\eta(d)- 1].
\end{equation}
The average of $I(m)$ over all outcomes,
\begin{equation}
 I=\sum_m p(m)\,I(m),
\end{equation}
is equal to the mutual information~\cite{FucPer96}
between the random variables $\{a\}$ and $\{m\}$,
\begin{equation}
   I=\sum_{m,a} p(m,a)\, \log_2\frac{p(m,a)}{p(m)\,p(a)}
\end{equation}
with $p(m,a)=p(m|a)\,p(a)$ because $p(a)$ is uniform.

Alternatively,
the state of the system can be estimated as a state $\ket{\varphi(m)}$
depending on the outcome $m$.
In the optimal estimation~\cite{Banasz01},
$\ket{\varphi(m)}$ is the eigenvector of $\hcs{\hat{M}_m}$
corresponding to its maximum eigenvalue.
The quality of the estimate is evaluated
by the estimation fidelity such that
\begin{equation}
 G(m) =\sum_a p(a|m)\,\bigl|\expv{\varphi(m)|\psi(a)}\bigr|^2.
\end{equation}
As was found for $I(m)$,
$G(m)$ also quantifies the amount of information
provided by the outcome $m$ of the measurement $\{\hat{M}_m\}$
[cf. Eq.~(\ref{eq:defInformation})]
and is explicitly written in terms of
the singular values of $\hat{M}_m$ as~\cite{Terash15}
\begin{equation}
 G(m) =\frac{1}{d+1}\left(\frac{\sigma_m^2
                +\lambda_{m,\max}^2}{\sigma_m^2}\right),
\label{eq:estimation}
\end{equation}
where $\lambda_{m,\max}$ is the maximum singular value of $\hat{M}_m$.
Note that $G(m)$ satisfies
\begin{equation}
  \frac{1}{d}\le G(m) \le \frac{2}{d+1}.
\end{equation}
The average of $G(m)$ over all outcomes,
\begin{equation}
  G=\sum_m p(m)\,G(m),
\end{equation}
becomes the mean estimation fidelity discussed in Ref.~\cite{Banasz01}
because
\begin{equation}
 p(m)=\frac{\sigma_m^2}{d}, \qquad \sum_m \sigma_m^2=d,
\end{equation}
even though $G(m)$ was not derived in Ref.~\cite{Banasz01}.
Note that $G$ can be derived from $G(m)$; however,
$G(m)$ cannot be derived from $G$.
That is,
$G(m)$ characterizes the measurement $\{\hat{M}_m\}$ 
in more detail than $G$.

Next, the degree of disturbance
caused by the measurement is quantified.
When the measurement $\{\hat{M}_m\}$ yields an outcome $m$,
the state of the system changes from $\ket{\psi(a)}$ to $\ket{\psi(m,a)}$,
as given in Eq.~(\ref{eq:postMeasurement}).
The size of this state change is evaluated
by the operation fidelity such that
\begin{equation}
 F(m) =\sum_a p(a|m)\bigl|\expv{\psi(a)|\psi(m,a)}\bigr|^2.
\label{eq:defFidelity}
\end{equation}
$F(m)$ quantifies the degree of disturbance caused
when the measurement $\{\hat{M}_m\}$ yields the outcome $m$
and is explicitly written in terms of
the singular values of $\hat{M}_m$ as~\cite{Terash15}
\begin{equation}
    F(m) =\frac{1}{d+1}\left(\frac{\sigma_m^2
                +\tau_m^2}{\sigma_m^2}\right),
\label{eq:fidelity}
\end{equation}
where
\begin{equation}
 \tau_m =\sum_{i}\lambda_{mi}.
\end{equation}
Note that $F(m)$ satisfies
\begin{equation}
 \frac{2}{d+1}\le F(m) \le 1.
\end{equation}
Similar to $G(m)$,
the average of $F(m)$ over all outcomes,
\begin{equation}
  F=\sum_m p(m)\,F(m),
\end{equation}
becomes the mean operation fidelity discussed in Ref.~\cite{Banasz01},
even though $F(m)$ was not derived in Ref.~\cite{Banasz01}.

In addition to the size of the state change,
the reversibility of the state change can also be regarded as
a measure of the disturbance.
Even though $\ket{\psi(a)}$ and $\ket{\psi(m,a)}$ are unknown,
this state change is physically reversible
if $\hat{M}_m$ has a bounded left
inverse $\hat{M}_m^{-1}$~\cite{UeImNa96,Ueda97}.
To recover $\ket{\psi(a)}$,
a second measurement called a reversing measurement
is made on $\ket{\psi(m,a)}$.
The reversing measurement is described by
another set of measurement operators $\{\hat{R}_\mu^{(m)}\}$ that satisfy
\begin{equation}
\sum_\mu\hat{R}^{(m)\dagger}_\mu\hat{R}^{(m)}_\mu=\hat{I},
\end{equation}
and, moreover, $\hat{R}^{(m)}_{\mu_0}\propto \hat{M}_m^{-1}$
for a particular $\mu=\mu_0$,
where $\mu$ denotes the outcome of the reversing measurement.
When the reversing measurement yields the preferred outcome $\mu_0$,
the state of the system reverts to $\ket{\psi(a)}$
via the state change caused by the reversing measurement
because $\hat{R}_{\mu_0}^{(m)}\hat{M}_m\propto\hat{I}$.
For the optimal reversing measurement~\cite{KoaUed99},
the probability of recovery is given by
\begin{equation}
 R(m,a)=\frac{\lambda_{m,\min}^2}{p(m|a)},
\end{equation}
where $\lambda_{m,\min}$ is the minimum singular value of $\hat{M}_m$.
The reversibility of the state change is then evaluated by this
maximum successful probability as
\begin{equation}
  R(m) = \sum_a p(a|m)\,R(m,a).
\end{equation}
As was found for $F(m)$,
$R(m)$ also quantifies the degree of disturbance
caused when the measurement $\{\hat{M}_m\}$
yields the outcome $m$ [cf. Eq.~(\ref{eq:defFidelity})]
and is explicitly written in terms of
the singular values of $\hat{M}_m$ as~\cite{Terash15}
\begin{equation}
  R(m) =d\left(\frac{\lambda_{m,\min}^2 }{\sigma_m^2}\right).
\label{eq:reversibility}
\end{equation}
Note that $R(m)$ satisfies
\begin{equation}
 0\le R(m) \le 1.
\end{equation}
The average of $R(m)$ over all outcomes,
\begin{equation}
 R=\sum_m p(m)\,R(m),
\end{equation}
is the degree of physical reversibility of a measurement
discussed in Ref.~\cite{KoaUed99},
whose explicit form in terms of the singular values is given
in Ref.~\cite{CheLee12},
even though $R(m)$ was not derived in Ref.~\cite{CheLee12}.

Therefore, the information and disturbance for a single outcome $m$
are obtained as functions of the singular values of $\hat{M}_m$:
$I(m)$ and $G(m)$ for the information and
$F(m)$ and $R(m)$ for the disturbance.
Note that they are invariant
under the interchange of any pair of singular values,
\begin{equation}
\lambda_{mi} \longleftrightarrow \lambda_{mj}
 \quad \text{for any $(i,j)$},
\label{eq:interchange}
\end{equation}
and under rescaling of all the singular values,
\begin{equation}
 \lambda_{mi} \longrightarrow  c\lambda_{mi}
 \quad \text{for all $i$},
\label{eq:rescaling}
\end{equation}
by a constant $c$~\cite{Terash15}.
By contrast, the probability for the outcome $m$, $p(m)=\sigma_m^2/d$,
is invariant under the interchange
but is not invariant under the rescaling.

As an important example, consider $\hat{M}^{(d)}_{k,l}(\lambda)$,
which is defined as a measurement operator
whose singular values are
\begin{equation}
  \underbrace{1,1,\ldots,1}_{k},
  \underbrace{\lambda,\lambda,\ldots,\lambda}_{l},
  \underbrace{0,0,\ldots,0}_{d-k-l}
\end{equation}
with $0\le\lambda\le1$.
Even though the information and disturbance for $\hat{M}^{(d)}_{k,l}(\lambda)$
can be calculated from
Eqs.~(\ref{eq:information}), (\ref{eq:estimation}),
(\ref{eq:fidelity}), and (\ref{eq:reversibility}),
calculating $I(m)$ is not straightforward
due to the degeneracy of the singular values.
By taking the limit $\lambda_{mi}\to\lambda_{mk}$,
$I(m)$ is found to be~\cite{Terash15}
\begin{align}
 I(m) &=\log_2d
  -\frac{1}{\ln2}\Bigl[\eta(d)- 1\Bigr]-
     \log_2\left(k+\lambda^2\right) \notag\\
&{}+\frac{1}{k+\lambda^2}\left[
\frac{\lambda^{2(k+1)}\log_2\lambda^2}
   {(\lambda^2-1)^k}
-\sum_{n=0}^{k-1}\frac{a^{(k+1)}_n}{(\lambda^2-1)^{k-n}}
\right]
\end{align}
for $\hat{M}^{(d)}_{k,1}(\lambda)$ and
\begin{align}
 I(m) &=\log_2d
  -\frac{1}{\ln2}\Bigl[\eta(d)- 1\Bigr] \notag\\
 &{}\quad-\log_2\left(1+l\lambda^2\right)-\frac{1}{1+l\lambda^2}
  \sum_{n=0}^{l-1}\frac{c^{(l+1)}_n(\lambda)}{(1-\lambda^2)^{l-n}}
\end{align}
for $\hat{M}^{(d)}_{1,l}(\lambda)$,
where $\{a^{(j)}_n\}$ and $\{c^{(j)}_n(\lambda)\}$ are given by
\begin{equation}
a^{(j)}_n =\frac{1}{\ln2}\binom{j}{n}
       \Bigl[\eta(j)- \eta(j-n)\Bigr],
\end{equation}
\begin{equation}
c^{(j)}_n(\lambda) = \lambda^{2(j-n)}
   \left[\binom{j}{n}\log_2 \lambda^2+a^{(j)}_n\right].
\end{equation}
Similarly, $\hat{P}^{(d)}_{r}$ is defined as
a projective measurement operator of rank $r$.
Note that $\hat{M}^{(d)}_{k,l}(0)=\hat{P}^{(d)}_{k}$,
$\hat{M}^{(d)}_{k,l}(1)=\hat{P}^{(d)}_{k+l}$,
and $\hat{P}^{(d)}_{d}=\hat{I}$.
For $\hat{P}^{(d)}_{r}$, $I(m)$ is found to be~\cite{Terash11b}
\begin{equation}
  I(m)=\log_2\frac{d}{r}-\frac{1}{\ln2}\Bigl[\eta(d)- \eta(r)\Bigr].
\end{equation}

\section{\label{sec:region}Allowed Region}
Next, we plot the information and disturbance
for various measurement operators on a plane.
In particular,
an allowed region for information versus disturbance
can be shown on the plane by plotting
all physically possible measurement operators; that is,
by varying every singular value over the range of $0\le\lambda_{mi}\le1$.
It is easy to do this
for $I(m)$, $G(m)$, $F(m)$, and $R(m)$
because they contain only a definite number of bounded parameters,
i.e., $d$ singular values,
in contrast to $I$, $G$, $F$, and $R$.
Moreover,
from the interchange invariance in Eq.~(\ref{eq:interchange}),
measurement operators having the same singular values
up to ordering
correspond to the same point on the plane.
According to the rescaling invariance in Eq.~(\ref{eq:rescaling}),
$\hat{M}_m$ and $c\hat{M}_m$ correspond to the same point on the plane.

\begin{figure}
\begin{center}
\includegraphics[scale=0.52]{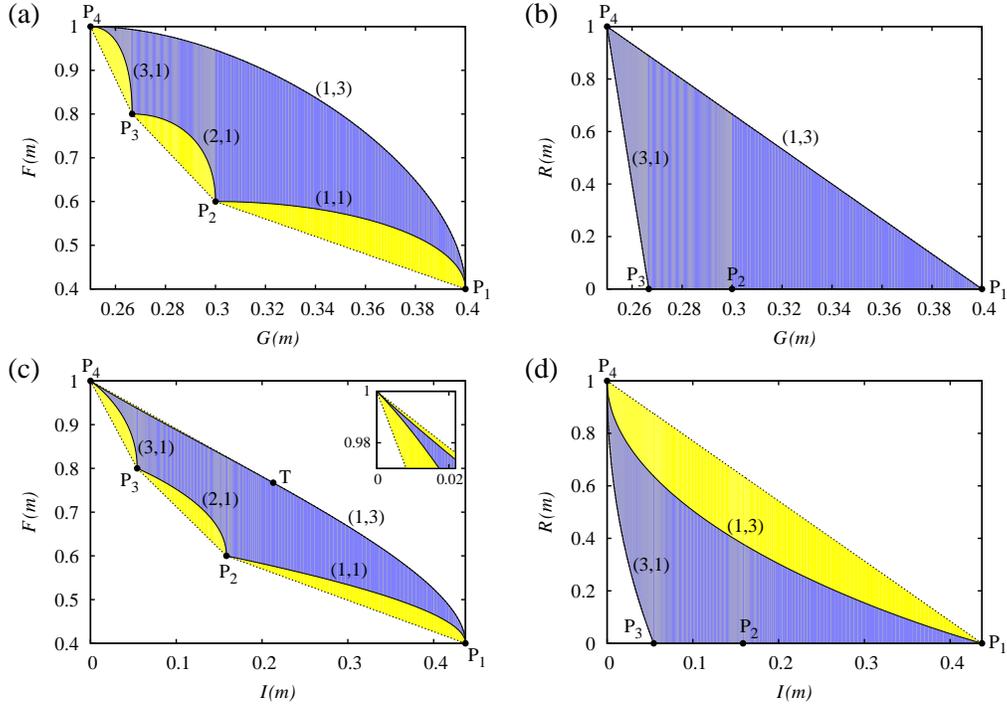}
\end{center}
\caption{\label{fig1}
Four allowed regions for information versus disturbance for $d=4$:
(a) estimation fidelity $G(m)$ versus operation fidelity $F(m)$,
(b) estimation fidelity $G(m)$ versus physical reversibility $R(m)$,
(c) information gain $I(m)$ versus operation fidelity $F(m)$, and
(d) information gain $I(m)$ versus physical reversibility $R(m)$.
In each panel,
the region pertaining to a single outcome is shown in blue (dark gray),
and the extended region obtained by averaging over all outcomes is shown
in yellow (light gray).
}
\end{figure}

\begin{figure}
\begin{center}
\includegraphics[scale=0.52]{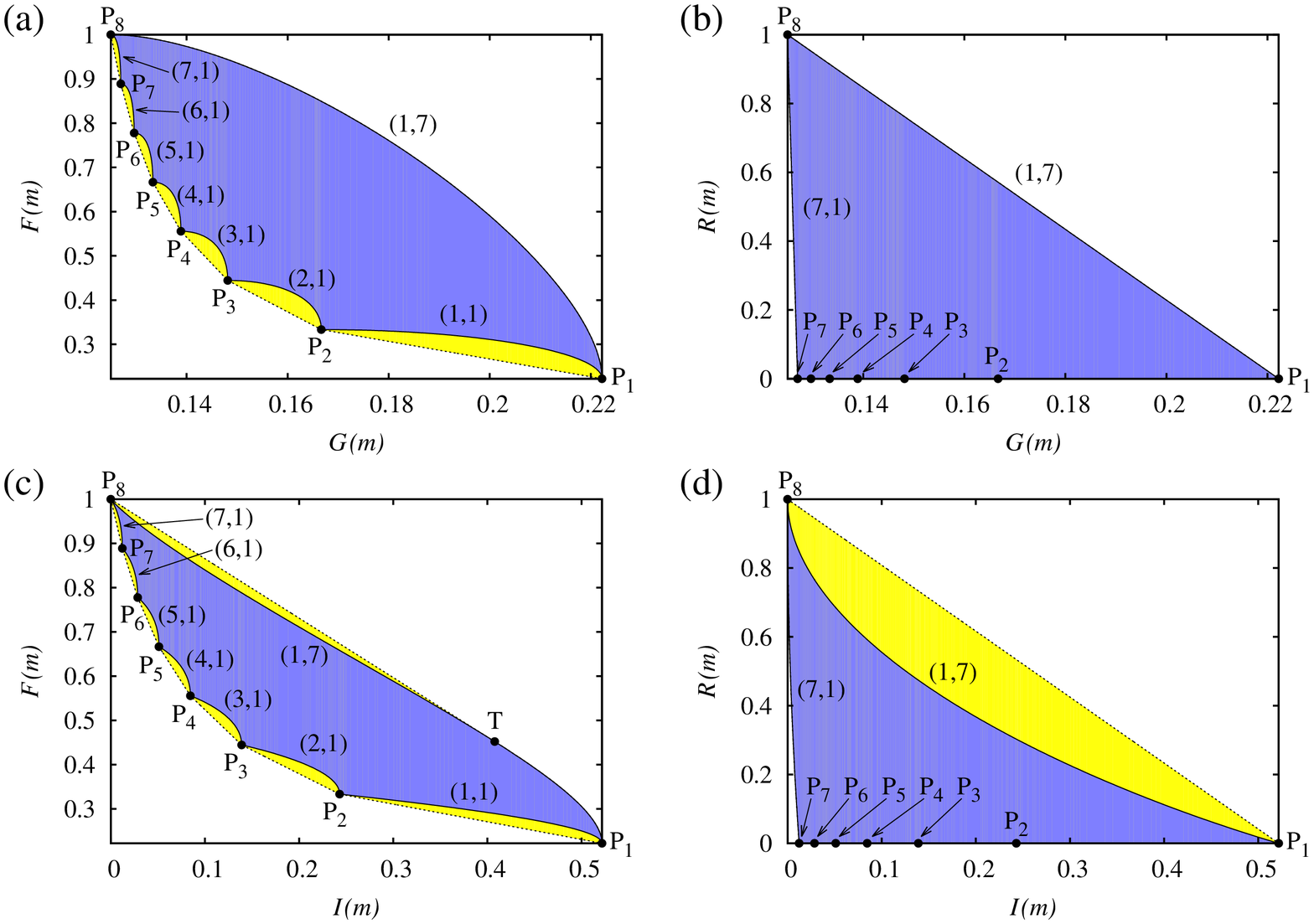}
\end{center}
\caption{\label{fig2}
Four allowed regions for information versus disturbance for $d=8$:
(a) estimation fidelity $G(m)$ versus operation fidelity $F(m)$,
(b) estimation fidelity $G(m)$ versus physical reversibility $R(m)$,
(c) information gain $I(m)$ versus operation fidelity $F(m)$, and
(d) information gain $I(m)$ versus physical reversibility $R(m)$.
In each panel,
the region pertaining to a single outcome is shown in blue (dark gray),
and the extended region obtained by averaging over all outcomes is shown
in yellow (light gray).
}
\end{figure}

Figure~\ref{fig1}(a) shows
the allowed region for $G(m)$ versus $F(m)$
when $d=4$ in blue (dark gray).
In the figure,
$\mathrm{P}_r$ and $(k,l)$ represent
the point corresponding to $c\hat{P}^{(d)}_{r}$ and
the line corresponding to $c\hat{M}^{(d)}_{k,l}(\lambda)$
with $0\le\lambda\le1$, respectively.
The upper boundary consists of one curved line $(1,d-1)$
connecting $\mathrm{P}_1$ and $\mathrm{P}_d$
as $\lambda$ varies from $0$ to $1$,
whereas the lower boundary consists of $d-1$ curved lines $(k,1)$
connecting $\mathrm{P}_k$ to $\mathrm{P}_{k+1}$
for $k=1,2,\ldots,d-1$.
Conversely, Fig.~\ref{fig1}(b) shows
the allowed region for $G(m)$ versus $R(m)$ when $d=4$
in blue (dark gray).
In this case,
both the upper and lower boundaries consist of one straight line:
$(1,d-1)$ for the upper boundary and
$(d-1,1)$ for the lower boundary.
Similarly,
Figs.~\ref{fig1}(c) and \ref{fig1}(d) show
the allowed region for $I(m)$ versus $F(m)$
and for $I(m)$ versus $R(m)$, respectively.
The measurement operators corresponding to the upper and lower boundaries
are the same as for $G(m)$, even though the lines have different shapes.
Figure~\ref{fig2} shows
the allowed regions when $d=8$ in blue (dark gray).

The above boundaries, $(1,d-1)$ and $(k,1)$, were first confirmed by
brute-force numerical calculations where every singular value was
varied by steps of
$\Delta \lambda_{mi}=0.01$ for $d=2,3,\ldots,6$
and $\Delta \lambda_{mi}=0.02$ for $d=7,8$.
Moreover, for $G(m)$ versus $F(m)$
and for $G(m)$ versus $R(m)$,
the boundaries can analytically be proven
to be the true boundaries for arbitrary $d$ (see Appendix \ref{sec:proof}).
Unfortunately, however,
for $I(m)$ versus $F(m)$ and for $I(m)$ versus $R(m)$,
proving that the boundaries are the true boundaries
is difficult analytically.
Nevertheless, they can be shown to satisfy
the necessary conditions for the true boundaries
using the Karush--Kuhn--Tucker (KKT) conditions~\cite{Avriel03},
which generalize the method of Lagrange multipliers
to handle inequality constraints in mathematical optimization.
For example, to find the lower boundary for $I(m)$ versus $F(m)$,
consider minimizing $I(m)$ subject to $F(m)=F_0$ and $\lambda_{mi}\ge0$
($i=1,2,\ldots,d$).
Then, $\hat{M}^{(d)}_{k,1}(\lambda)$ satisfies
a necessary condition for a local minimum,
that is, for a Lagrange function
\begin{equation}
L_F=I(m)- \alpha_F \left[F(m)-F_0\right]-\sum_i\beta_i\lambda_{mi},
\end{equation}
$\hat{M}^{(d)}_{k,1}(\lambda)$
satisfies $\partial L_F/\partial\lambda_{mi}=0$ with
KKT multipliers $\alpha_F$ and $\{\beta_i\}$ such that 
$\beta_i\ge0$ and $\beta_i\lambda_{mi}=0$ for all $i$
and has $\lambda=\lambda_0$ such that $F(m)=F_0$
if $(k+1)/(d+1)\le F_0\le(k+2)/(d+1)$.
These mathematical optimizations are explained
in Appendix \ref{sec:optimization}.

\section{\label{sec:average}Average over Outcomes}
Here, the regions that are allowed for
the information and disturbance averaged over all possible outcomes
are discussed:
$I$ and $G$ for the information and $F$ and $R$ for the disturbance.
Unfortunately, it is difficult to show the allowed regions
directly from their explicit forms written in terms of the singular values
because the number of singular values contained in them
is not definite due to the indefinite number of outcomes.
Note that
there are no physical limitations on the number of outcomes.

Instead, we show the allowed regions
using the following analogy with the center of mass.
In the measurement $\{\hat{M}_m\}$,
each measurement operator $\hat{M}_m$ corresponds
to a point $\mathrm{R}_m$
in the allowed region pertaining to a single outcome
with weight $p(m)$.
This situation can be viewed as a set of particles,
each with a mass $p(m)$
located at a point $\mathrm{R}_m$.
The center of mass of these particles then indicates
the average information and disturbance of the measurement.
Conversely, for an arbitrary set of particles
located in the allowed region pertaining to a single outcome,
an equivalent measurement satisfying Eq.~(\ref{eq:completeness})
can be constructed
by rescaling and duplicating the measurement operators,
as shown in Appendix \ref{sec:construction}.
For example, for $d=4$, two particles with the same mass $1/2$
located at $\mathrm{P}_1$ and $\mathrm{P}_4$ in Fig.~\ref{fig1}
can be simulated by a measurement with five outcomes
whose measurement operators are
\begin{equation}
  \hat{M}_m
   =\begin{cases}
      \frac{1}{\sqrt{2}}\,\ket{m}\bra{m}
         & \mbox{($m=1,2,3,4$)} \\[10pt]
      \frac{1}{\sqrt{2}}\,\hat{I}
         & \mbox{($m=5$)}.
    \end{cases}
\end{equation}
Therefore, the allowed region for
the average information and disturbance can be shown
by considering the center of mass of
all possible sets of particles.
Note that the center of mass may be located outside the region
where the particles are situated,
which means that the allowed region is extended
by averaging over the outcomes.
The resultant region is the convex hull of the original region.

The regions extended by averaging
are shown in Fig.~\ref{fig1} in yellow (light gray).
As shown in Fig.~\ref{fig1}(a),
the lower boundary for $G$ versus $F$ is extended
to the straight lines between
$\mathrm{P}_k$ and $\mathrm{P}_{k+1}$ for $k=1,2,\ldots,d-1$,
whereas the upper boundary is not extended due to its convexity.
By contrast, as shown in Fig.~\ref{fig1}(b),
the boundaries for $G$ versus $R$ are not extended at all.
Meanwhile,
as shown in Fig.~\ref{fig1}(c),
the lower boundary for $I$ versus $F$ is extended as in the case of $G$
and, moreover,
the upper boundary is extended a little higher when $d\ge3$
because the line $(1,d-1)$ has a slight dent near $\mathrm{P}_d$.
In fact,
an analytic calculation of $\hat{M}^{(d)}_{1,d-1}(\lambda)$
shows that
\begin{equation}
 \frac{d^2F(m)}{dI(m)^2}>0
\end{equation}
near $\mathrm{P}_d$ when $d\ge3$.
The upper boundary is therefore extended
to the tangent line drawn from $\mathrm{P}_d$ to the line $(1,d-1)$
between $\mathrm{P}_d$ and the point of tangency $\mathrm{T}$.
As shown in Fig.~\ref{fig1}(d), 
the upper boundary for $I$ versus $R$ is extended to
the straight line between $\mathrm{P}_1$ and $\mathrm{P}_d$,
whereas the lower boundary is not extended.
The case of $d=8$ is shown in Fig.~\ref{fig2}.

\begin{figure}
\begin{center}
\includegraphics[scale=0.6]{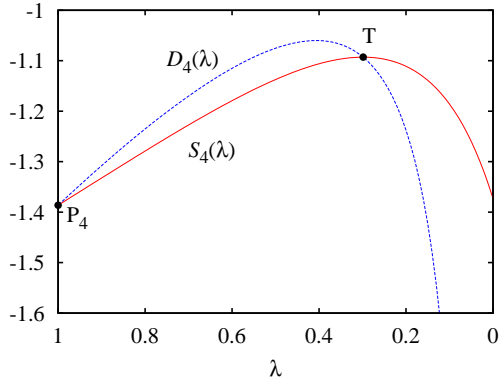}
\end{center}
\caption{\label{fig3}
Two line slopes for $d=4$.
$D_4(\lambda)$ is the slope of the tangent line to the line $(1,3)$
at a point $\mathrm{Q}$,
and $S_4(\lambda)$ is the slope of the straight line
from $\mathrm{P}_4$ to $\mathrm{Q}$.
Note that the horizontal axis is reversed.}
\end{figure}
\begin{figure}
\begin{center}
\includegraphics[scale=0.6]{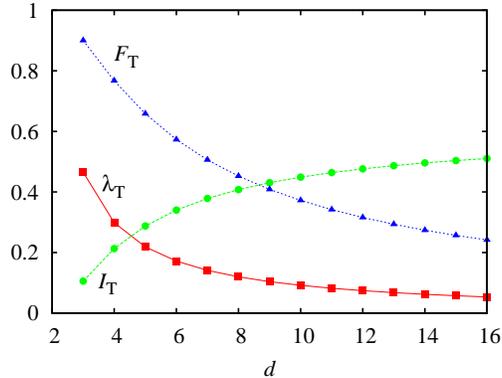}
\end{center}
\caption{\label{fig4}
Singular value $\lambda_\mathrm{T}$,
information $I_\mathrm{T}$, and fidelity $F_\mathrm{T}$
at the point of tangency $\mathrm{T}$ for various $d$.}
\end{figure}

%
To find the point $\mathrm{T}$ on the upper boundary for $I$ versus $F$,
two line slopes are defined as functions of $\lambda$:
the slope of the tangent line to the line $(1,d-1)$
at the point $\mathrm{Q}$ corresponding to $\hat{M}^{(d)}_{1,d-1}(\lambda)$,
\begin{equation}
  D_d(\lambda)=\frac{dF(m)}{dI(m)},
\label{eq:slopeD}
\end{equation}
and the slope of the straight line from
$\mathrm{P}_d$ to $\mathrm{Q}$,
\begin{equation}
  S_d(\lambda)=\frac{F(m)-1}{I(m)}.
\label{eq:slopeS}
\end{equation}
These functions are shown for $d=4$ in Fig.~\ref{fig3}.
Using $\lambda_\mathrm{T}$ such that
\begin{equation}
D_d(\lambda_\mathrm{T})=S_d(\lambda_\mathrm{T}),
\label{eq:lambdaT}
\end{equation}
the measurement operator corresponding to $\mathrm{T}$
can be written as
$\hat{M}^{(d)}_{1,d-1}(\lambda_\mathrm{T})$.
In Fig.~\ref{fig4},
$\lambda_\mathrm{T}$ is shown
with $I(m)$ and $F(m)$ at $\mathrm{T}$,
denoted by $I_\mathrm{T}$ and $F_\mathrm{T}$, respectively,
for various $d$.
When $d=4$, $\mathrm{T}$ in Fig.~\ref{fig1}(c)
corresponds to $\hat{M}^{(4)}_{1,3}(0.299)$
and the upper boundary for $I$ versus $F$
moves up between $\mathrm{P}_4$ and $\mathrm{T}$,
at most by $3.5\times 10^{-3}$.
This extension of the upper boundary 
becomes larger as $d$ increases.
For example,
when $d=8$, $\mathrm{T}$ in Fig.~\ref{fig2}(c)
corresponds to $\hat{M}^{(8)}_{1,7}(0.120)$
and the upper boundary
moves up at most by $2.6\times 10^{-2}$.
Interestingly, $\hat{M}^{(d)}_{1,d-1}(\lambda_\mathrm{T})$ is
the most efficient measurement operator
in terms of the ratio of information gain to fidelity loss~\cite{Terash15},
\begin{equation}
E_F(m)=\frac{I(m)}{1-F(m)}.
\end{equation}

The upper boundary for $G$ versus $F$
and that for $G$ versus $R$ are equivalent to the inequalities of
Banaszek~\cite{Banasz01} and Cheong and Lee~\cite{CheLee12}, respectively,
where the averages are explicitly calculated using $p(m)=\sigma_m^2/d$.
However, to our knowledge, this is the first derivation of
the other two upper and four lower boundaries.
The lower boundaries are less important than the upper boundaries
in quantum information and can be violated by non-ideal measurements,
which have classical noise yielding mixed post-measurement states,
or by non-optimal estimations, which assume
unwise observers making incorrect choices
for $\ket{\varphi(m)}$ in $G(m)$.
Nevertheless, for the foundations of quantum mechanics,
it is worth deriving both the upper and lower boundaries
for ideal measurements with optimal estimation
to examine the intrinsic nature and power of quantum measurements.

\begin{figure}
\begin{center}
\includegraphics[scale=0.52]{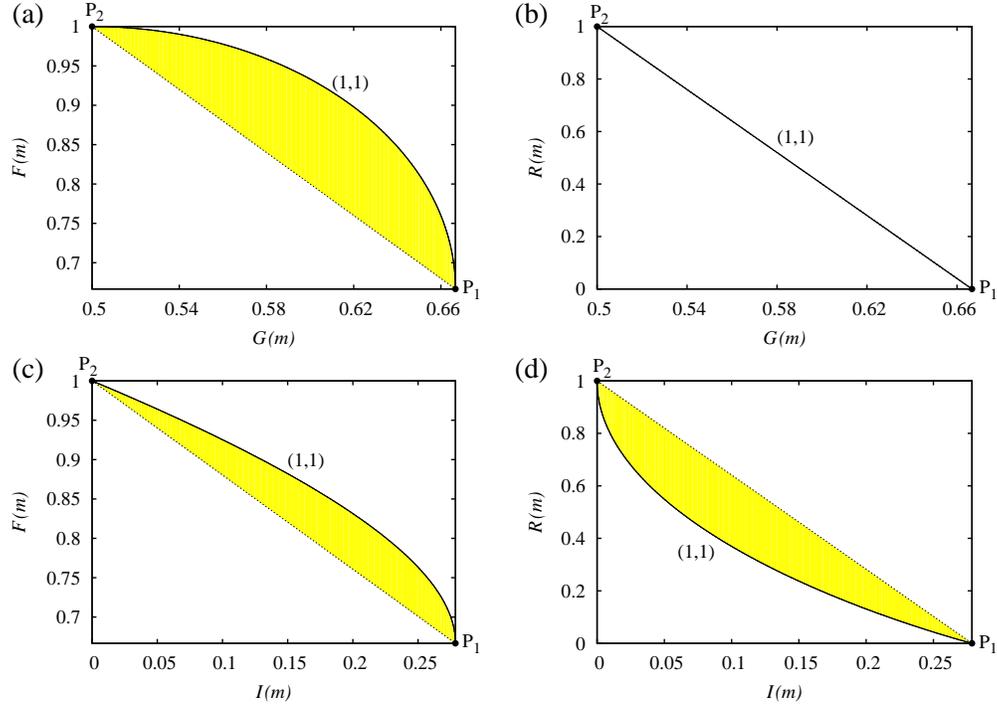}
\end{center}
\caption{\label{fig5}
Four allowed regions for information versus disturbance for $d=2$:
(a) estimation fidelity $G(m)$ versus operation fidelity $F(m)$,
(b) estimation fidelity $G(m)$ versus physical reversibility $R(m)$,
(c) information gain $I(m)$ versus operation fidelity $F(m)$, and
(d) information gain $I(m)$ versus physical reversibility $R(m)$.
In each panel,
the region pertaining to a single outcome is
just the solid line denoted by $(1,1)$ and
the extended region obtained by averaging over all outcomes is shown
in yellow (light gray).
}
\end{figure}
\begin{figure}
\begin{center}
\includegraphics[scale=0.6]{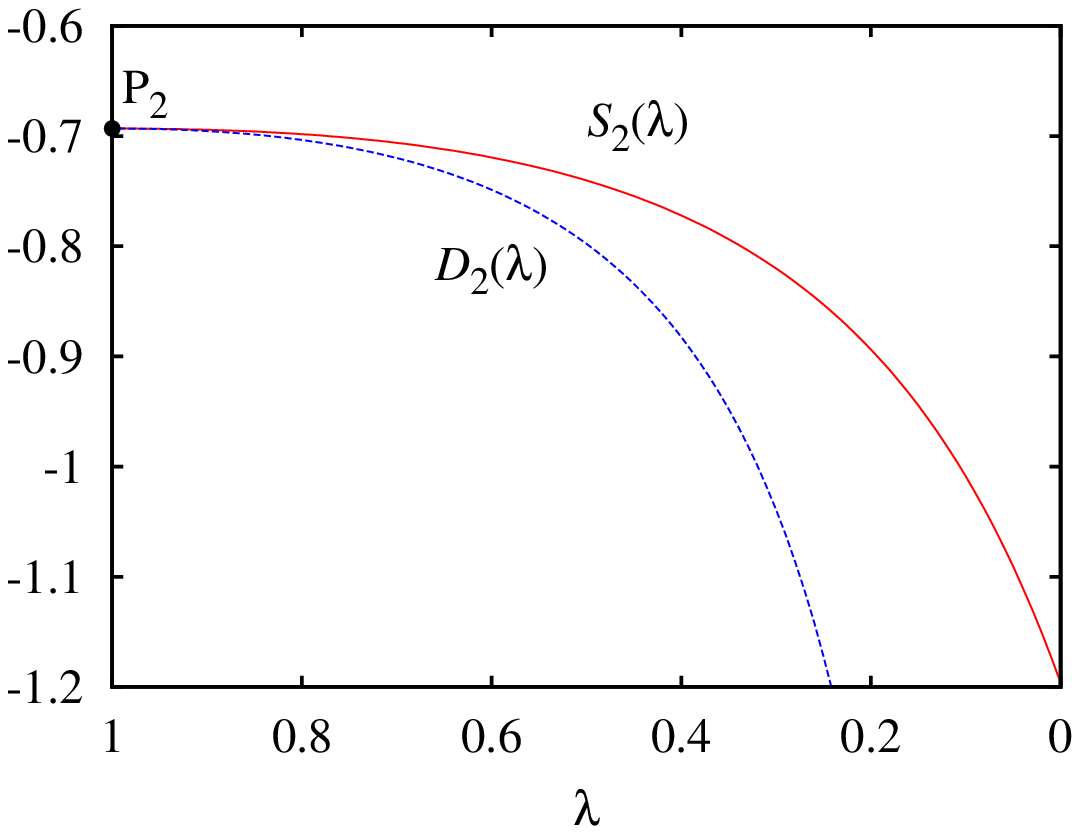}
\end{center}
\caption{\label{fig6}
Two line slopes for $d=2$.
$D_2(\lambda)$ is the slope of the tangent line to the line $(1,1)$
at a point $\mathrm{Q}$,
and $S_2(\lambda)$ is the slope of the straight line
from $\mathrm{P}_2$ to $\mathrm{Q}$.
Note that the horizontal axis is reversed.}
\end{figure}

%
The case of $d=2$ is a special case, where
the regions extended by averaging are the main parts of
the allowed regions, as shown in Fig.~\ref{fig5}.
In this case, the allowed regions pertaining to a single outcome
shrink to the line $(1,1)$
because a measurement operator can be represented
by a single parameter via the rescaling invariance
in Eq.~(\ref{eq:rescaling})~\cite{Terash11}.
Moreover, the line $(1,1)$ in Fig.~\ref{fig5}(c) has
no dent unlike the case of $d\ge3$.
In fact,
it can be shown for $\hat{M}^{(2)}_{1,1}(\lambda)$ that
\begin{equation}
 \frac{d^2F(m)}{dI(m)^2}<0
\end{equation}
near $\mathrm{P}_2$.
The point $\mathrm{T}$ does not exist on the line $(1,1)$
because the slopes $D_2(\lambda)$ and $S_2(\lambda)$
in Eqs.~(\ref{eq:slopeD}) and (\ref{eq:slopeS})
do not become equal to each other
except for $\lambda=1$, as shown in Fig.~\ref{fig6}.

\section{\label{sec:optimal}Optimal Measurement}
Finally, we discuss the optimal measurements
saturating the upper bounds on the information
for a given disturbance.
The upper bounds are denoted by the upper boundaries
of the allowed regions for the average information and disturbance.
Therefore, according to the analogy with the center of mass,
a measurement is optimal for an information--disturbance pair
if it is equivalent to a set of particles
whose center of mass is on the upper boundary
for that information--disturbance pair.
The optimal measurements are different for
the four types of information--disturbance pairs
because the upper boundaries have different shapes
on the four information--disturbance planes,
as shown in Fig.~\ref{fig1}.

The conditions for the optimal measurements are as follows.
A measurement $\{\hat{M}_m\}$ is optimal for $G$ versus $F$
if all $\hat{M}_m$'s correspond
to an identical point on the line $(1,d-1)$
because the upper boundary for $G$ versus $F$ is
the convex curve $(1,d-1)$,
as shown in Fig.~\ref{fig1}(a),
whereas it is optimal for $G$ versus $R$
if every $\hat{M}_m$ corresponds to a point on the line $(1,d-1)$
because the upper boundary for $G$ versus $R$ is
the straight line $(1,d-1)$,
as shown in Fig.~\ref{fig1}(b).
These conditions are
equivalent to those in Refs.~\cite{Banasz01,CheLee12}.
Similarly, when $d\ge3$,
a measurement $\{\hat{M}_m\}$ is optimal for $I$ versus $F$
if all $\hat{M}_m$'s correspond
to an identical point between $\mathrm{T}$ and $\mathrm{P}_1$
on the line $(1,d-1)$ or
if every $\hat{M}_m$ corresponds to either $\mathrm{P}_d$ or $\mathrm{T}$
because the upper boundary for $I$ versus $F$ is the union of
the convex curve $(1,d-1)$
between $\mathrm{T}$ and $\mathrm{P}_1$ and
the straight line between $\mathrm{P}_d$ and $\mathrm{T}$,
as shown in Fig.~\ref{fig1}(c).
However, when $d=2$,
the condition to be optimal for $I$ versus $F$
is the same as that for $G$ versus $F$
because the upper boundary is just the convex curve $(1,d-1)$,
as shown in Fig.~\ref{fig5}(c).
Conversely,
a measurement $\{\hat{M}_m\}$ is optimal for $I$ versus $R$
if every $\hat{M}_m$ corresponds to either $\mathrm{P}_d$ or $\mathrm{P}_1$
because the upper boundary for $I$ versus $R$
is the straight line between $\mathrm{P}_d$ and $\mathrm{P}_1$,
as shown in Fig.~\ref{fig1}(d).

\begin{figure}
\begin{center}
\includegraphics[scale=0.42]{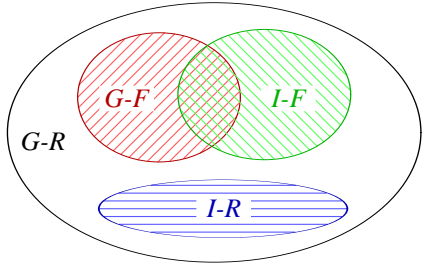}
\end{center}
\caption{\label{fig7}
Four conditions for optimal measurements.
For example,
the set $G$-$F$ represents all measurements
that are optimal for $G$ versus $F$.}
\end{figure}

Interestingly, an optimal measurement for $G$ versus $F$ is
not necessarily optimal for $I$ versus $F$
and an optimal measurement for $G$ versus $R$ is
not necessarily optimal for $I$ versus $R$.
The relationships between the four conditions are illustrated
in Fig.~\ref{fig7},
excluding the strongest measurement,
where all the measurement operators correspond to $\mathrm{P}_1$,
and the weakest measurement,
where all the measurement operators correspond to $\mathrm{P}_d$;
these two measurements satisfy all four conditions.

As a specific example,
consider a measurement $\{\hat{M}_{m}^{(d)}(\lambda)\}$
with $d$ outcomes, $m=1,2,\ldots,d$,
where $\hat{M}_{m}^{(d)}(\lambda)$ is defined by
\begin{equation}
  \hat{M}_{m}^{(d)}(\lambda)\equiv\frac{1}{\sqrt{1+(d-1)\lambda^2}}
  \left(\ket{m}\bra{m}+\sum_{i\neq m} \lambda \ket{i}\bra{i}\right)
\end{equation}
with $0<\lambda<1$.
For a given $\lambda$,
all $\hat{M}_{m}^{(d)}(\lambda)$'s
correspond to an identical point on the line $(1,d-1)$
in the four information--disturbance planes
because they are equivalent to $\hat{M}^{(d)}_{1,d-1}(\lambda)$
via the interchange and rescaling invariances
in Eqs.~(\ref{eq:interchange}) and (\ref{eq:rescaling}).
The corresponding point on the line $(1,d-1)$
indicates the average information and disturbance
of $\{\hat{M}_{m}^{(d)}(\lambda)\}$.
The measurement $\{\hat{M}_{m}^{(d)}(\lambda)\}$ is optimal
both for $G$ versus $F$ and for $G$ versus $R$ for arbitrary $\lambda$
because the line $(1,d-1)$ is equal to
the upper boundary, as shown in
Figs.~\ref{fig1}(a) and \ref{fig1}(b).

However, the measurement $\{\hat{M}_{m}^{(d)}(\lambda)\}$
is not necessarily optimal for $I$ versus $F$
because only a part of the line $(1,d-1)$ is
equal to the upper boundary when $d\ge3$,
as shown in Fig.~\ref{fig1}(c).
It is optimal for $I$ versus $F$
only if $\lambda\le \lambda_\mathrm{T}$,
with $\lambda_\mathrm{T}$
being defined by Eq.~(\ref{eq:lambdaT}).
Note that
$\hat{M}_{m}^{(d)}(\lambda_\mathrm{T})$ corresponds to
$\mathrm{T}$ on the S-shaped curve $(1,d-1)$.
If $\lambda> \lambda_\mathrm{T}$,
$\hat{M}_{m}^{(d)}(\lambda)$ corresponds to a point
on the concave part between $\mathrm{P}_d$ and $\mathrm{T}$
of the line $(1,d-1)$,
where the upper boundary is equal to
the straight line between $\mathrm{P}_d$ and $\mathrm{T}$.
This means that $\{\hat{M}_{m}^{(d)}(\lambda)\}$
is not optimal for $I$ versus $F$ if $\lambda> \lambda_\mathrm{T}$
or equivalently if $F>F_\mathrm{T}$.
The optimal measurement for this case can easily be constructed
from the analogy with the center of mass by considering two particles:
one located at $\mathrm{T}$ with mass $q$ and
the other located at $\mathrm{P}_d$ with mass $1-q$.
According to Appendix \ref{sec:construction},
the optimal measurement has $d+1$ outcomes
whose measurement operators are
\begin{equation}
  \hat{M}_m
   =\begin{cases}
      \sqrt{q}\,\hat{M}_{m}^{(d)}(\lambda_\mathrm{T})
         & \mbox{($m=1,2,\ldots,d$)} \\[10pt]
      \sqrt{1-q}\,\hat{I}
         & \mbox{($m=d+1$)},
    \end{cases}
\label{eq:optIF}
\end{equation}
where $q=(1-F)/\left(1-F_\mathrm{T}\right)$
for a given $F$.
The average information and disturbance of this measurement
are then indicated by a point
on the straight line between $\mathrm{P}_d$ and $\mathrm{T}$
equal to a part of the upper boundary.
By contrast, when $d=2$,
$\{\hat{M}_{m}^{(2)}(\lambda)\}$
is optimal for $I$ versus $F$ for arbitrary $\lambda$
because the line $(1,1)$ is equal to
the upper boundary, as shown in Fig.~\ref{fig5}(c).

Conversely,
the measurement $\{\hat{M}_{m}^{(d)}(\lambda)\}$ is not optimal
for $I$ versus $R$ for any $\lambda$
because the line $(1,d-1)$ is not equal to
the upper boundary at all, as shown in Fig.~\ref{fig1}(d).
In this case, the upper boundary is the straight line
between $\mathrm{P}_d$ and $\mathrm{P}_1$.
Therefore,
the optimal measurement for $I$ versus $R$ can be constructed
from the analogy with the center of mass by considering two particles:
one located at $\mathrm{P}_1$ with mass $q$
and the other located at $\mathrm{P}_d$ with mass $1-q$.
This has $d+1$ outcomes whose measurement operators are
\begin{equation}
  \hat{M}_m
   =\begin{cases}
      \sqrt{q}\,\ket{m}\bra{m}
         & \mbox{($m=1,2,\ldots,d$)} \\[10pt]
      \sqrt{1-q}\,\hat{I}
         & \mbox{($m=d+1$)},
    \end{cases}
\label{eq:optIR}
\end{equation}
where $q=1-R$ for a given $R$.
The average information and disturbance of this measurement
are indicated by a point
on the straight line between $\mathrm{P}_d$ and $\mathrm{P}_1$
equal to the upper boundary.

Of course,
the measurements given in Eqs.~(\ref{eq:optIF}) and (\ref{eq:optIR}) are
also optimal for $G$ versus $R$ for arbitrary $q$.
Even though their measurement operators
correspond to different points on the line $(1,d-1)$,
the point indicating the average values
is still on the line $(1,d-1)$ equal to the upper boundary
because the line $(1,d-1)$ is straight,
as shown in Fig.~\ref{fig1}(b).
However, except for $q=0$ or $1$,
the measurement in Eq.~(\ref{eq:optIF}) is optimal neither 
for $G$ versus $F$ nor for $I$ versus $R$
and the measurement in Eq.~(\ref{eq:optIR}) is optimal neither
for $G$ versus $F$ nor for $I$ versus $F$.

\section{\label{sec:summary}Summary}
In summary, we have shown the allowed regions
for information versus disturbance
for quantum measurements of completely unknown states.
The information and disturbances
pertaining to a single outcome are quantified
using the singular values of the measurement operator
and are plotted on four types of information--disturbance planes
to show the allowed regions pertaining to a single outcome.
The allowed regions for the average values are also discussed
via an analogy with the center of mass.
These regions explicitly give not only the upper bounds
but also the lower bounds on the information for a given disturbance
together with the optimal measurements saturating the upper bounds.
Consequently,
our results broaden our perspective of quantum measurements
and provide a useful tool for quantum information processing
and communication.

\appendix
\section*{Appendix}

\section{\label{sec:proof}Proof of Boundaries}
Here, the proofs of the boundaries are outlined
for $G(m)$ versus $F(m)$ and for $G(m)$ versus $R(m)$.
To prove the upper and lower boundaries for $G(m)$ versus $F(m)$,
consider maximizing and minimizing $F(m)$ for a given $G(m)$.
Using the interchange and rescaling invariances
in Eqs.~(\ref{eq:interchange}) and (\ref{eq:rescaling}),
the singular values are assumed to be sorted in descending order,
$\lambda_{m1}\ge\lambda_{m2}\ge\cdots\ge\lambda_{md}$,
and normalized such that $\sigma_m^2=1$.
Then, the problems are simplified
to maximizing and minimizing $\sum_{i=2}^{d} \lambda_{mi}$
subject to $\sum_{i=2}^{d} \lambda_{mi}^2=1-\lambda_{m1}^2$
and $0\le\lambda_{mi}\le\lambda_{m1}$ for a given $\lambda_{m1}$
from Eqs.~(\ref{eq:estimation}) and (\ref{eq:fidelity}).
The maximum is achieved when
$\lambda_{m2}=\lambda_{m3}=\cdots=\lambda_{md}
=\sqrt{(1-\lambda_{m1}^2)/(d-1)}$.
The corresponding singular values are proportional to
those of $\hat{M}^{(d)}_{1,d-1}(\lambda)$
with $\lambda=\sqrt{(1-\lambda_{m1}^2)/(d-1)}/\lambda_{m1}$.
Therefore,
the line $(1,d-1)$ is the upper boundary for $G(m)$ versus $F(m)$.

Conversely,
the minimum is achieved
when $\lambda_{m2}=\sqrt{1-\lambda_{m1}^2}$ and the others are $0$
if $\lambda_{m1}\ge 1/\sqrt{2}$.
Because these singular values are proportional to
those of $\hat{M}^{(d)}_{1,1}(\lambda)$
with $\lambda=\sqrt{1-\lambda_{m1}^2}/\lambda_{m1}$,
the line $(1,1)$ is the lower boundary for $G(m)$ versus $F(m)$
if $G(m)\ge 3/(2d+2)$.
However, if $\lambda_{m1}< 1/\sqrt{2}$,
they do not satisfy $\lambda_{m2}\le\lambda_{m1}$
because $\sqrt{1-\lambda_{m1}^2}>\lambda_{m1}$.
Therefore, in this case, let $\lambda_{m2}=\lambda_{m1}$ and
consider minimizing $\sum_{i=3}^{d} \lambda_{mi}$
subject to $\sum_{i=3}^{d} \lambda_{mi}^2=1-2\lambda_{m1}^2$
and $0\le\lambda_{mi}\le\lambda_{m1}$ for a given $\lambda_{m1}$.
If $\lambda_{m1}\ge 1/\sqrt{3}$,
the minimum is achieved when
$\lambda_{m3}=\sqrt{1-2\lambda_{m1}^2}$ and the others are $0$.
Because these singular values are proportional to
those of $\hat{M}^{(d)}_{2,1}(\lambda)$
with $\lambda=\sqrt{1-2\lambda_{m1}^2}/\lambda_{m1}$,
the line $(2,1)$ is the lower boundary for $G(m)$ versus $F(m)$
if $4/(3d+3) \le G(m)<3/(2d+2)$.
By repeating similar minimizations for $\lambda_{m1}< 1/\sqrt{3}$,
the lines $(k,1)$ with $k=1,2,\ldots,d-1$ are shown to be
the lower boundaries for $G(m)$ versus $F(m)$.

Similarly, to prove the upper and lower boundaries
for $G(m)$ versus $R(m)$,
consider maximizing and minimizing $R(m)$ for a given $G(m)$.
Via the descending ordering and the normalization $\sigma_m^2=1$,
the problems are simplified
to maximizing and minimizing $\lambda_{md}$
subject to $\sum_{i=2}^{d} \lambda_{mi}^2=1-\lambda_{m1}^2$
and $0\le\lambda_{mi}\le\lambda_{m1}$ for a given $\lambda_{m1}$
from Eqs.~(\ref{eq:estimation}) and (\ref{eq:reversibility}).
As in the case of $F(m)$,
the maximum is achieved
when $\lambda_{m2}=\lambda_{m3}=\cdots=\lambda_{md}$.
This result shows that
the line $(1,d-1)$ is the upper boundary for $G(m)$ versus $R(m)$.
Conversely,
the minimum is achieved when $\lambda_{md}=0$
if $\lambda_{m1}\ge1/\sqrt{d-1}$.
That is,
$R(m)=0$ is the lower boundary for $G(m)$ versus $R(m)$
if $G(m)\ge d/(d^2-1)$.
However, if $\lambda_{m1}<1/\sqrt{d-1}$,
$\lambda_{md}$ cannot be $0$ to satisfy $\sigma_m^2=1$
because $\lambda_{mi}\le\lambda_{m1}$.
In this case, the minimum is achieved
when $\lambda_{md}=\sqrt{1-(d-1)\lambda_{m1}^2}$
and the others are $\lambda_{m1}$.
These singular values are proportional to
those of $\hat{M}^{(d)}_{d-1,1}(\lambda)$
with $\lambda=\sqrt{1-(d-1)\lambda_{m1}^2}/\lambda_{m1}$.
This result shows that
the line $(d-1,1)$ is the lower boundary for $G(m)$ versus $R(m)$
if $G(m)< d/(d^2-1)$.

\section{\label{sec:optimization}Mathematical Optimization}
Here, the mathematical optimizations of the information
for a given disturbance are outlined
for $I(m)$ versus $F(m)$ and for $I(m)$ versus $R(m)$
based on the method of Lagrange multipliers
and its generalization known as
the Karush--Kuhn--Tucker (KKT) conditions~\cite{Avriel03}.
Consider maximizing $I(m)$ subject to $F(m)=F_0$
using a Lagrange function $L^F=-I(m)-\alpha^F [F(m)-F_0]$
with a multiplier $\alpha^F$.
To use the method of Lagrange multipliers,
the derivatives of $I(m)$ and $F(m)$
with respect to $\lambda_{mi}$ should be calculated.
From the rescaling invariance in Eq.~(\ref{eq:rescaling}),
the derivatives of $I(m)$ satisfy
\begin{equation}
  \sum_i\lambda_{mi}\frac{\partial I(m)}{\partial\lambda_{mi}}=0
\end{equation}
according to Euler's homogeneous function theorem.
Using this equation and
the interchange invariance in Eq.~(\ref{eq:interchange}),
the derivatives of $I(m)$ for $\hat{M}^{(d)}_{k,l}(\lambda)$
can be written as
\begin{equation}
  \frac{\partial I(m)}{\partial\lambda_{mi}}
   \equiv \begin{cases}
       \mathcal{I}^{(d)}_{k,l}(\lambda)
               & \mbox{($1\le i \le k$)} \\[10pt]
       -\frac{k}{l\lambda}\mathcal{I}^{(d)}_{k,l}(\lambda)
               & \mbox{($k+1\le i \le k+l$)} \\[10pt]
       0
               & \mbox{($k+l+1\le i \le d$)}
    \end{cases}
\end{equation}
with $\mathcal{I}^{(d)}_{k,l}(\lambda)\ge 0$,
where the third case is $0$ because
$I(m)$ is a function of $\{\lambda_{mi}^2\}$.
Similarly, the derivatives of $F(m)$
for $\hat{M}^{(d)}_{k,l}(\lambda)$ can be written as
\begin{equation}
  \frac{\partial F(m)}{\partial\lambda_{mi}}
   \equiv \begin{cases}
      \mathcal{F}^{(d)}_{k,l}(\lambda)
               & \mbox{($1\le i \le k$)} \\[10pt]
      -\frac{k}{l\lambda}\mathcal{F}^{(d)}_{k,l}(\lambda)
               & \mbox{($k+1\le i \le k+l$)} \\[10pt]
      \widetilde{\mathcal{F}}^{(d)}_{k,l}(\lambda)
               & \mbox{($k+l+1\le i \le d$)}
    \end{cases}
\end{equation}
with $\mathcal{F}^{(d)}_{k,l}(\lambda)\le 0$ and
$\widetilde{\mathcal{F}}^{(d)}_{k,l}(\lambda)> 0$.
These derivatives show that $\hat{M}^{(d)}_{1,d-1}(\lambda)$
satisfies $\partial L^F/\partial\lambda_{mi}=0$ for all $i$
with a multiplier of
$\alpha^F=-\mathcal{I}^{(d)}_{1,d-1}(\lambda)
/\mathcal{F}^{(d)}_{1,d-1}(\lambda)$.
Moreover, there exists a parameter $\lambda_0$
such that $F(m)$ for $\hat{M}^{(d)}_{1,d-1}(\lambda_0)$ is equal to $F_0$.
That is,
$\hat{M}^{(d)}_{1,d-1}(\lambda_0)$ satisfies
a necessary condition for a local maximum
according to the method of Lagrange multipliers.
This result implies that the line $(1,d-1)$ is the upper boundary
for $I(m)$ versus $F(m)$.

Conversely, consider minimizing $I(m)$ subject to $F(m)=F_0$
and $\lambda_{mi}\ge0$ ($i=1,2,\ldots,d$).
The inequality constraints are indispensable in this case
because the solutions are
on the boundary of the parameter space, $\lambda_{mi}=0$.
To handle these inequality constraints,
the KKT conditions are applied
using a Lagrange function
$L_F=I(m)- \alpha_F \left[F(m)-F_0\right]-\sum_i\beta_i\lambda_{mi}$
with multipliers $\alpha_F$ and $\{\beta_i\}$.
Then, $\hat{M}^{(d)}_{k,1}(\lambda)$ satisfies
$\partial L_F/\partial\lambda_{mi}=0$ for all $i$
with multipliers
$\alpha_F=\mathcal{I}^{(d)}_{k,1}(\lambda)
/\mathcal{F}^{(d)}_{k,1}(\lambda)$
and
\begin{equation}
  \beta_i
   =\begin{cases}
       0  & \mbox{($1\le i \le k+1$)} \\[10pt]
       -\alpha_F\widetilde{\mathcal{F}}^{(d)}_{k,1}(\lambda)
               & \mbox{($k+2\le i \le d$)}.
    \end{cases}
\end{equation}
In addition, these $\{\beta_i\}$ satisfy
the requirements as multipliers for the inequality constraints,
$\beta_i\ge 0$ and $\beta_i\lambda_{mi}=0$, for all $i$.
There exists a parameter $\lambda_0$ such that
$F(m)$ for $\hat{M}^{(d)}_{k,1}(\lambda_0)$ is equal to $F_0$
if $(k+1)/(d+1)\le F_0\le(k+2)/(d+1)$.
That is,
$\hat{M}^{(d)}_{k,1}(\lambda_0)$
satisfies a necessary condition for a local minimum
according to the KKT conditions.
This result implies that the line $(k,1)$ is the lower boundary
for $I(m)$ versus $F(m)$.

Similarly,
letting $\lambda_{m,\min}=\lambda_{md}$,
consider maximizing $I(m)$ subject to $R(m)=R_0$
and $\lambda_{mi}-\lambda_{md}\ge0$ ($i=1,2,\ldots,d-1$)
using a Lagrange function
$L^R=-I(m)-\alpha^R [R(m)-R_0]-\sum_{i}\gamma_i (\lambda_{mi}-\lambda_{md})$
with multipliers $\alpha^R$ and $\{\gamma_i\}$.
The derivatives of $R(m)$
for $\hat{M}^{(d)}_{k,l}(\lambda)$ can be written when $k+l=d$ such that
\begin{equation}
  \frac{\partial R(m)}{\partial\lambda_{mi}}
   \equiv\begin{cases}
      \mathcal{R}^{(d)}_{k,l}(\lambda)
               & \mbox{($1\le i \le k$)} \\[10pt]
      -\frac{k}{l\lambda}\mathcal{R}^{(d)}_{k,l}(\lambda)
          -\frac{1-l\delta_{i,d}}{l}
        \widetilde{\mathcal{R}}^{(d)}_{k,l}(\lambda)
               & \mbox{($k+1\le i \le d$)}
    \end{cases}
\end{equation}
with $\mathcal{R}^{(d)}_{k,l}(\lambda)\le 0$ and
$\widetilde{\mathcal{R}}^{(d)}_{k,l}(\lambda)\ge 0$.
Then, $\hat{M}^{(d)}_{1,d-1}(\lambda)$
satisfies $\partial L^R/\partial\lambda_{mi}=0$ for all $i$
with multipliers
$\alpha^R=-\mathcal{I}^{(d)}_{1,d-1}(\lambda)
/\mathcal{R}^{(d)}_{1,d-1}(\lambda)$
and
\begin{equation}
  \gamma_i
   =\begin{cases}
       0  & \mbox{($i=1$)} \\[10pt]
       \frac{1}{d-1}\alpha^R\widetilde{\mathcal{R}}^{(d)}_{1,d-1}(\lambda) 
          & \mbox{($2\le i \le d-1$)}
    \end{cases}
\end{equation}
satisfying $\gamma_i\ge 0$ and
$\gamma_i(\lambda_{mi}-\lambda_{md})=0$ for all $i$.
Moreover, there exists a parameter $\lambda_0$ such that
$R(m)$ for $\hat{M}^{(d)}_{1,d-1}(\lambda_0)$ is equal to $R_0$.
According to the KKT conditions, $\hat{M}^{(d)}_{1,d-1}(\lambda_0)$
satisfies a necessary condition for a local maximum
implying that the line $(1,d-1)$ is the upper boundary
for $I(m)$ versus $R(m)$.
Conversely, consider minimizing $I(m)$ subject to $R(m)=R_0$
using a Lagrange function $L_R=I(m)-\alpha_R [R(m)-R_0]$
with a multiplier $\alpha_R$.
Then, $\hat{M}^{(d)}_{d-1,1}(\lambda)$
satisfies $\partial L_R/\partial\lambda_{mi}=0$ for all $i$
with the multiplier
$\alpha_R=\mathcal{I}^{(d)}_{d-1,1}(\lambda)
/\mathcal{R}^{(d)}_{d-1,1}(\lambda)$
and there exists a parameter $\lambda_0$ such that
$R(m)$ for $\hat{M}^{(d)}_{d-1,1}(\lambda_0)$ is equal to $R_0$.
According to the method of Lagrange multipliers,
$\hat{M}^{(d)}_{d-1,1}(\lambda_0)$
satisfies a necessary condition for a local minimum
implying that the line $(d-1,1)$ is the lower boundary
for $I(m)$ versus $R(m)$.

\section{\label{sec:construction}Construction of Equivalent Measurement}
Here,
the general construction of an equivalent measurement
is presented for an arbitrary set of particles
located in the allowed region pertaining to a single outcome.
The construction is not trivial
because a measurement operator
not only corresponds to a point
but also gives the weight at that point.
Moreover, the measurement operators must satisfy
Eq.~(\ref{eq:completeness}).

Consider a set of particles, where
each particle $n$ has a mass $q_n$ and is located at a point $\mathrm{R}_n$
in the allowed region pertaining to a single outcome.
Without a loss of generality,
the total mass can be assumed to be $\sum_n q_n=1$.
By definition,
there exists a measurement operator $\hat{M}_n$
with singular values $\{\lambda_{ni}\}$
that corresponds to the point $\mathrm{R}_n$.
In general,
its weight $p(n)=\sigma_n^2/d$
is not equal to the mass $q_n$.
However, the weight can be adjusted by
rescaling and duplicating $\hat{M}_n$.
That is, for a particle $n$,
$d$ measurement operators are introduced such that
\begin{equation}
\hat{M}_{ns}\equiv \sqrt{\frac{q_n}{\sigma_n^2}}
\sum_{i} \lambda_{ni} \ket{c_s(i)}\bra{c_s(i)}
\end{equation}
with $s=0,1,\ldots,d-1$,
where $c_s(i)\equiv(i-1+s \mod d)+1$ performs
the cyclic permutation of $\{\ket{i}\}$.
These measurement operators correspond to the same point $\mathrm{R}_n$
from the interchange invariance in Eq.~(\ref{eq:interchange}),
giving the same weight $q_n/d$.
Note that the weight is not invariant
under rescaling of the singular values in Eq.~(\ref{eq:rescaling}).
The total weight of the $d$ measurement operators
is then equal to the mass $q_n$ as desired.
Moreover, such measurement operators for all the particles
satisfy Eq.~(\ref{eq:completeness}) such that
$\sum_{n,s}\hcs{\hat{M}_{ns}}=\sum_n q_n\hat{I} = \hat{I}$
when regarding a pair of indices $(n,s)$ as an outcome $m$.
Therefore, $\{\hat{M}_{ns}\}$ is a measurement
equivalent to the set of particles.

In this construction,
one particle corresponds to $d$ outcomes,
even though the number of outcomes can be
reduced when some singular values are degenerate.
As a result,
it suffices to consider measurements
having at most $2d$ outcomes
to study the allowed regions for the average values
because for any point in the region there exists
a set of two particles whose center of mass is located at that point.


\begin{thebibliography}{10}

\bibitem{NieChu00}
M.~A. Nielsen and I.~L. Chuang, {\em Quantum Computation and Quantum
  Information} (Cambridge University Press, Cambridge, 2000).

\bibitem{BenBra84}
C.~H. Bennett and G. Brassard,  in {\em Proceedings of IEEE International
  Conference on Computers, Systems and Signal Processing, Bangalore, India}
  (IEEE, New York, 1984), pp.\ 175--179.

\bibitem{Ekert91}
A.~K. Ekert, Phys. Rev. Lett. {\bf 67},  661  (1991).

\bibitem{Bennet92}
C.~H. Bennett, Phys. Rev. Lett. {\bf 68},  3121  (1992).

\bibitem{BeBrMe92}
C.~H. Bennett, G. Brassard, and N.~D. Mermin, Phys. Rev. Lett. {\bf 68},  557
  (1992).

\bibitem{FucPer96}
C.~A. Fuchs and A. Peres, Phys. Rev. A {\bf 53},  2038  (1996).

\bibitem{Banasz01}
K. Banaszek, Phys. Rev. Lett. {\bf 86},  1366  (2001).

\bibitem{FucJac01}
C.~A. Fuchs and K. Jacobs, Phys. Rev. A {\bf 63},  062305  (2001).

\bibitem{BanDev01}
K. Banaszek and I. Devetak, Phys. Rev. A {\bf 64},  052307  (2001).

\bibitem{Barnum02}
H. Barnum, arXiv:quant-ph/0205155.

\bibitem{DArian03}
G.~M. D'Ariano, Fortschr. Phys. {\bf 51},  318  (2003).

\bibitem{Ozawa04}
M. Ozawa, Ann. Phys. (NY) {\bf 311},  350  (2004).

\bibitem{GenPar05}
M.~G. Genoni and M.~G.~A. Paris, Phys. Rev. A {\bf 71},  052307  (2005).

\bibitem{MiFiFi05}
L. Mi{\v{s}}ta, Jr., J. Fiur{\'a}{\v{s}}ek, and R. Filip,
Phys. Rev. A {\bf 72},  012311  (2005).

\bibitem{Maccon06}
L. Maccone, Phys. Rev. A {\bf 73},  042307  (2006).

\bibitem{Sacchi06}
M.~F. Sacchi, Phys. Rev. Lett. {\bf 96},  220502  (2006).

\bibitem{BusSac06}
F. Buscemi and M.~F. Sacchi, Phys. Rev. A {\bf 74},  052320  (2006).

\bibitem{Banasz06}
K. Banaszek, Open Syst. Inf. Dyn. {\bf 13},  1  (2006).

\bibitem{BuHaHo07}
F. Buscemi, M. Hayashi, and M. Horodecki, Phys. Rev. Lett. {\bf 100},  210504
  (2008).

\bibitem{RenFan14}
X.-J. Ren and H. Fan, J. Phys. A: Math. Theor. {\bf 47},  305302  (2014).

\bibitem{FGNZ15}
L. Fan, W. Ge, H. Nha, and M.~S. Zubairy, Phys. Rev. A {\bf 92},  022114
  (2015).

\bibitem{ShKuUe16}
T. Shitara, Y. Kuramochi, and M. Ueda, Phys. Rev. A {\bf 93},  032134  (2016).

\bibitem{Terash10}
H. Terashima, Phys. Rev. A {\bf 83},  032111  (2011).

\bibitem{Terash11}
H. Terashima, Phys. Rev. A {\bf 83},  032114  (2011).

\bibitem{CheLee12}
Y.~W. Cheong and S.-W. Lee, Phys. Rev. Lett. {\bf 109},  150402  (2012).

\bibitem{Terash15}
H. Terashima, Phys. Rev. A {\bf 93},  022104  (2016).

\bibitem{UedKit92}
M. Ueda and M. Kitagawa, Phys. Rev. Lett. {\bf 68},  3424  (1992).

\bibitem{UeImNa96}
M. Ueda, N. Imoto, and H. Nagaoka, Phys. Rev. A {\bf 53},  3808  (1996).

\bibitem{Ueda97}
M. Ueda, in {\em Frontiers in Quantum Physics: Proceedings of the
  International Conference on Frontiers in Quantum Physics,
  Kuala Lumpur, Malaysia, 1997},
  edited by S.~C. Lim, R. Abd-Shukor, and K.~H. Kwek
  (Springer, Singapore, 1998), pp.\ 136--144.

\bibitem{SRDFM06}
F. Sciarrino, M. Ricci, F. De~Martini, R. Filip, and L. Mi{\v{s}}ta, Jr.,
Phys. Rev. Lett. {\bf 96},  020408  (2006).

\bibitem{BaChKi08}
S.-Y. Baek, Y.~W. Cheong, and Y.-H. Kim, Phys. Rev. A {\bf 77},  060308(R)
  (2008).

\bibitem{CZXTLX14}
G. Chen, Y. Zou, X.-Y. Xu, J.-S. Tang, Y.-L. Li, J.-S. Xu, Y.-J. Han,
C.-F. Li, G.-C. Guo, H.-Q. Ni, Y. Yu, M.-F. Li, G.-W. Zha, Z.-C. Niu,
and Y. Kedem, Phys. Rev. X {\bf 4},  021043  (2014).

\bibitem{LRHLK14}
H.-T. Lim, Y.-S. Ra, K.-H. Hong, S.-W. Lee, and Y.-H. Kim,
Phys. Rev. Lett. {\bf 113},  020504  (2014).

\bibitem{KoaUed99}
M. Koashi and M. Ueda, Phys. Rev. Lett. {\bf 82},  2598  (1999).

\bibitem{Royer94}
A. Royer, Phys. Rev. Lett. {\bf 73},  913  (1994);
{\bf 74}, 1040(E) (1995).

\bibitem{NieCav97}
M.~A. Nielsen and C.~M. Caves, Phys. Rev. A {\bf 55},  2547  (1997).

\bibitem{TerUed07b}
H. Terashima and M. Ueda, Phys. Rev. A {\bf 81},  012110  (2010).

\bibitem{Terash11b}
H. Terashima, Phys. Rev. A {\bf 85},  022124  (2012).

\bibitem{Avriel03}
M. Avriel, {\em Nonlinear Programming: Analysis and Methods} (Dover
  Publications, New York, 2003).

\end{thebibliography}

\end{document}